\documentclass[%
 reprint,
nofootinbib,
 amsmath,amssymb,
 aps,
]{revtex4-2}

\usepackage{graphicx}
\usepackage{dcolumn}
\usepackage{bm}
\usepackage{siunitx}
\usepackage{todonotes}
\usepackage{comment}
\usepackage{hyperref}

\definecolor{DARKMAGENTA}{HTML}{8b008b}

\begin{document}

\preprint{APS/123-QED}

\title{Can primordial parity violation explain the observed cosmic birefringence?}

\author{Tomohiro Fujita}
\affiliation{Waseda Institute for Advanced Study, Shinjuku, Tokyo 169-8050, Japan}%
\affiliation{Research Center for the Early Universe, The University of Tokyo, Bunkyo, Tokyo 113-0033, Japan}%
\email{tomofuji@aoni.waseda.jp}
\author{Yuto Minami}%
\affiliation{%
Research Center for Nuclear Physics, Osaka University, Ibaraki, Osaka, 567-0047, Japan
}%
\author{Maresuke Shiraishi}%
\affiliation{%
School of General and Management Studies, Suwa University of Science, 5000-1 Toyohira, Chino, Nagano 391-0292, Japan}%
\author{Shuichiro Yokoyama}
\affiliation{Kobayashi Maskawa Institute, Nagoya University, 
Chikusa, Aichi 464-8602, Japan}
\affiliation{Kavli IPMU (WPI), UTIAS, The University of Tokyo, 
Kashiwa, Chiba 277-8583, Japan}


\begin{abstract}
Recently, the cross-correlation between $E$- and $B$-mode polarization of the cosmic microwave background (CMB), which is well explained by cosmic birefringence with rotation angle $\beta\approx \SI{0.3}{deg}$, has been found in CMB polarization data. 
We carefully investigate the possibility of explaining the observed $EB$ correlation by the primordial chiral gravitational waves (CGWs), which can be generated in the parity-violating theories in the primordial Universe.
We found that the CGWs scenario does not work due to the overproduction of the $BB$ auto-correlation which far exceeds the observed one by SPTPol and POLARBEAR.
\end{abstract}

\maketitle


\section{\label{sec:intro}Introduction}
Measurements of linear polarization patterns of the cosmic microwave background (CMB) have provided rich information of our Universe~\cite{Komatsu:2014ioa,Aghanim:2018eyx, POLARBEAR:2022dxa, Polarbear:2020lii, ACT:2020gnv, SPT:2019nip,Dutcher:2021vtw,BICEP:2021xfz,Ade:2021cyk}.
The CMB photons gained the linear polarization at the last scattering surface (LSS) at redshift $z\approx1100$.
In addition, the polarization pattern may have been affected by some interactions, 
which the CMB photons experienced during the travel to observers. 
The linear polarization patterns of CMB can be decomposed into two orthogonal components:
parity even $E$-mode and parity odd $B$-mode.
Having them, we can construct two parity even correlations, $EE$ and $BB$,
and a parity odd correlation, $EB$.
If the parity symmetry is conserved for the CMB photons, the $EB$ cross-correlation should vanish, and thus it is a good probe of parity-violating physics.

Recently, a hint of parity violation was measured in the $EB$ correlation of the CMB photons~\cite{Minami:2020odp,Diego-Palazuelos:2022dsq,Eskilt:2022cff},
where a parity-violating physics, so-called ``cosmic birefringence'' was assumed.
Cosmic birefringence is that the linear polarization plane of the CMB photons rotates by angle $\beta$ while they travel between the LSS and observers.
This rotation produces a non-zero $EB$ cross-correlation from $EE$ and $BB$ spectra generated at the LSS as $ C_\ell^{EB,\rm{obs}} = \sin(4\beta) (C_\ell^{EE} - C_\ell^{BB})/2$~\cite{Minami:2019ruj}.
The recent careful data analysis indicates $\beta\approx 0.3\,\deg$ with about 3$\sigma$ significance~\cite{Eskilt:2022cff}.
No significant dependence of $\beta$ on the CMB photon frequency was found~\cite{Eskilt:2022wav}.
To explain this cosmic birefringence, several models have been proposed~\cite{Fujita:2020ecn,Takahashi:2020tqv,Fung:2021wbz,Choi:2021aze,
Obata:2021nql,Gasparotto:2022uqo}.

However, the non-zero detection of the $EB$ correlation does not necessarily imply that this parity-violating signal was caused by cosmic birefringence.
It is important to consider whether an alternative explanation is possible or not.
Primordial chiral gravitational waves (CGWs) are known to produce $EB$ correlation before the LSS because their imbalance between right- and left-handed circular polarization mode breaks the parity symmetry~\cite{Lue:1998mq}.
A number of models, which generate CGWs with various spectrum shapes in the primordial Universe, have been studied~\cite{Alexander:2004wk,Takahashi:2009wc,Anber:2012du,Adshead:2012kp,Namba:2015gja,Obata:2016tmo,Dimastrogiovanni:2016fuu,Adshead:2017hnc,Fujita:2018ndp,Machado:2018nqk,McDonough:2018xzh,Iarygina:2021bxq}. 
It is apparently possible to reproduce the observed $EB$ spectrum $C_{\ell}^{EB,{\rm obs}}$ by considering CGWs with a suitable spectrum shape.

In this short paper, we investigate if CGWs can consistently explain the observed $EB$ spectrum.
To test the $EB$ spectrum induced by CGWs with the observed $EB$,
one needs to re-estimate the miscalibration angles of detectors,
which are 
simultaneously determined in the measurements of $\beta$~\cite{Minami:2020odp,Diego-Palazuelos:2022dsq,Eskilt:2022cff}.
Because CGWs produce not only the $EB$ spectrum but also the other spectra between $T,E, B$,
an appropriate likelihood function that includes all spectra produced by CGWs is required, which is a very complicated task. However, we have noticed that the induced $BB$ spectrum becomes much larger than the observed value in the CGW scenario.
Therefore, we adopted a strategy of focusing on the compatibility of the $EB$ and $BB$ spectra induced by CGWs which are tuned to mimic the cosmic birefringence signal of $\beta=0.3\,\deg$.
We do not specify the generation mechanism of CGWs but introduce a spectrum template of CGWs with a sufficient number of parameters to obtain the desired $EB$ spectrum. We shall show that such CGWs lead to the overproduction of the $BB$ spectrum.

This paper is organized as follows.
In Sec.~\ref{sec:PCGW}, we introduce our template of CGWs.
In Sec.~\ref{sec:result}, we tune the parameters of the CGW spectrum and obtain the $EB$ spectrum similar to the observed one. 
In Sec.~\ref{BB result}, we compute the $BB$ spectrum induced by the CGW spectrum and show that it exceeds the observed value.
Sec.~\ref{sec:conclusion} is devoted to summary and discussion.

\section{\label{sec:PCGW}Chiral Gravitational Waves}

Chiral gravitational waves (CGWs), which violate the parity symmetry, produce $EB$ cross-correlation in the CMB polarization anisotropy if they are generated before the recombination era. 
In this section, we introduce our parameterization of the primordial spectrum of CGWs and illustrate how they induce CMB polarization correlations.

CGWs can be generated in the early Universe in various models, which predict diverse CGW spectrum shapes~\cite{Alexander:2004wk,Takahashi:2009wc,Anber:2012du,Adshead:2012kp,Namba:2015gja,Obata:2016tmo,Dimastrogiovanni:2016fuu,Adshead:2017hnc,Fujita:2018ndp,Machado:2018nqk,McDonough:2018xzh,Iarygina:2021bxq}.
In this paper, however, we adopt a model-independent approach.
We consider fully chiral and log-normal spectra of primordial CGWs and superpose them with different heights and peak positions as
\begin{align}
\mathcal{P}_h^{L}(k) 
&\simeq 0,
\notag\\
\mathcal{P}_h^{R}(k) 
&=\mathcal{P}_\zeta \sum_i r_{i}
 \exp \left[-\frac{1}{2\sigma^2}\ln^2\left(\frac{k}{k_i}\right)\right],
\label{eq:Ph_template}
\end{align}
where $\mathcal{P}_h^{L/R}$ denotes the dimensionless power spectrum of the left/right-handed circular polarization modes of the primordial gravitational waves on super-horizon scales.
$\mathcal{P}_\zeta=2.2\times 10^{-9}$ is the curvature power spectrum on the CMB scale.
$r_i, k_i, \sigma$ parameterize the amplitude, peak scale, and width of $\mathcal{P}_h^{R}$, respectively.
Our template \eqref{eq:Ph_template} can accommodate a sufficiently large parameter space to try mimicking the observed $EB$ spectrum, though we only introduce the single parameter $\sigma$ to control the width for simplicity. It should be stressed that we do not propose Eq.~\eqref{eq:Ph_template} as a natural spectrum shape of CGWs. Instead, we will show that even such a highly fine-tuned spectrum fails to explain the observation, and hence it is even harder for more realistic spectra.
Note that we consider the parity violation with $\mathcal{P}_h^{R} \gg \mathcal{P}_h^{L}$ so that CGWs induce positive $EB$ cross-correlation, which is favored by the measurements~\cite{Minami:2020odp, Diego-Palazuelos:2022dsq,Eskilt:2022cff}.

Primordial CGWs produce all of the auto- and cross-correlations between the CMB temperature $T$ and linear polarization $E$ and $B$.
Among them, we focus on $EB$ and $BB$ angular power spectra in this paper.
The contributions from the CGWs to them are written as e.g.,~\cite{Pritchard:2004qp,Namba:2015gja,Thorne:2017jft}
\begin{align}\begin{split}
    C_{\ell}^{EB} &= 4\pi \int {\rm d} (\ln k)
    \left[\mathcal{P}_h^{L}(k)-\mathcal{P}_h^{R}(k)\right] \Delta^E_\ell(k) \Delta^B_\ell(k),\\
    C_{\ell}^{BB} &= 4\pi \int {\rm d} (\ln k)
    \left[\mathcal{P}_h^{L}(k)+\mathcal{P}_h^{R}(k)\right] \Delta^B_\ell(k) \Delta^B_\ell(k),
\label{eq:EBBB}
\end{split}\end{align}
where $\Delta^{E/B}_\ell(k)$ is the tensor transfer function of the CMB $E$/$B$-mode,
and the contributions from the scalar and vector 
modes are ignored. 
Since $C_{\ell}^{EB}$ and $C_{\ell}^{BB}$ are linear functions of $\mathcal{P}_h^{R}$, each term in our template \eqref{eq:Ph_template} independently contributes to them.

\section{\label{sec:result} Reproducing $EB$ spectrum}

In this section, we numerically compute the $EB$ spectrum, which is contributed by the CGWs parametrized in Eq.~\eqref{eq:Ph_template}, using a modified version of a publicly available Boltzmann code CAMB \cite{Lewis:1999bs,camb} where Eq.~\eqref{eq:EBBB} is implemented. 
We shall fix the parameters, $k_i, r_i$ and $\sigma$ in Eq.~\eqref{eq:Ph_template}, in order to maximally mimic the observed $EB$ spectra.

\begin{figure*}
    \centering
    \includegraphics[width=\linewidth]{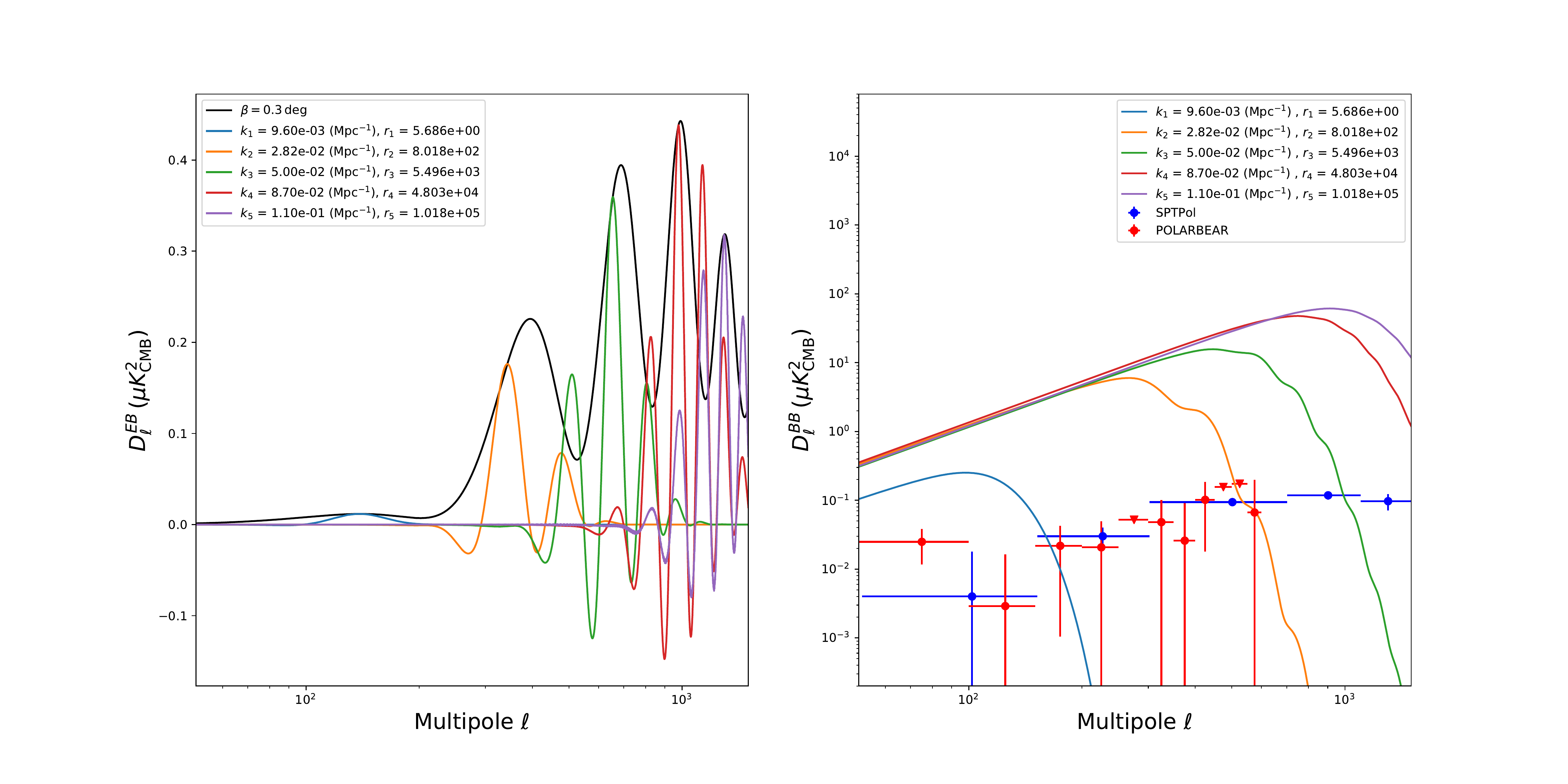}
    \caption{{\it (Left panel)} $EB$ angular power spectra $D_\ell^{EB}\equiv \ell(\ell+1) C_\ell^{EB}/(2\pi)$ against multipole $\ell$.
    The black line denotes $D_\ell^{EB}$ induced by cosmic birefringence with rotation angle of $\beta=0.3\,\deg$
    and we try reproducing it.
    The other colored lines represent individual $D_\ell^{EB}$ contributions produced by each term in chiral gravitational waves (CGWs) spectrum \eqref{eq:Ph_template} with the parameters tuned to match the peaks of the black line.
    {\it (Right panel)} 
    $BB$ angular power spectra $D_\ell^{BB}$ produced by the same CGWs as the left panel in the same color.
    Blue and red plots represent the actual $BB$ power spectra observed by SPTPol and POLARBEAR,
    which include foreground emissions, and any $D_\ell^{BB}$ prediction above them should be excluded.
    Circle dots and vertical bars show central values and $1\,\sigma~(68\,\%)$ uncertainties, respectively. 
    For the plots with negative center values, we show $95\%$ C.L. upper bound with inverted triangles. 
    }
    \label{fig:EBandBB}
\end{figure*}

In Fig.~\ref{fig:EBandBB}, we plot
$D_\ell^{XY} = {\ell(\ell+1)}  C_\ell^{XY}/{2\pi}~
(XY=EB, BB)$,
and the left panel is for the $EB$ spectrum and the right panel
shows the $BB$ spectrum.
In the left panel, 
the black solid line 
represents the target $EB$ spectrum inferred by the birefrigence angle reported by Refs.~\cite{Minami:2020odp,Diego-Palazuelos:2022dsq,Eskilt:2022cff}. One observes that the target $EB$ has 
five peaks up to $\ell \sim 1500$, which is the $\ell_\mathrm{max}$ used in the measurements of $\beta$~\cite{Minami:2020odp,Diego-Palazuelos:2022dsq,Eskilt:2022wav,Eskilt:2022cff},
that inherits from the intrinsic $EE$ spectrum, as cosmic birefringence induces, $ C_\ell^{EB,\rm{obs}} \simeq \sin(4\beta)\, C_\ell^{EE}/2$.
The colored lines 
denote the calculated $EB$ spectra produced by the CGWs. 
In mimicking the target $EB$ with CGWs, we introduce five terms in $\mathcal{P}_h^{R}$ to individually fit these peaks. 
To obtain narrow $EB$ peaks from the CGWs, we first set the width parameter to a small value $\sigma=0.2$, while we will vary it later.
Then, we search for the value of the peak scale $k_i$
and the peak amplitude $r_i$, 
to ensure that the calculated $EB$ spectrum is adjusted to the position and the height of each peak in the target $EB$.
In this manner, we determine five sets of parameters as
$\{k_i\,\SI{}{Mpc},\, r_i\} = \{\SI{9.60e-3},\, \SI{5.68e+0}\}, \{\SI{2.82e-2},\, \SI{8.02e2}\},$ $\{\SI{5.00e-2},\,\SI{ 5.50e+3}\}, \{\SI{8.70e-2},\,\SI{4.80e+04}\},$ $\{\SI{1.10e-1},\,\SI{1.02e+5}\}$.
One observes that these $EB$ spectra indueced by the CGWs show faster damped oscillations than the target one due to a distinctive feature of the tensor transfer function.
Note that their peak position and height do not completely coincide with the target peaks, because they are very sensitive to the free parameters and we did not pursue such a fine-tuning, which would not affect our conclusion. 

One might wonder if a similar multiple peak structure in the $EB$ spectrum could be reproduced by the tensor transfer functions without superposing five different log-normal spectra of CGWs. However, a single wide CGW spectrum leads to a $EB$ spectrum that has only a couple of peaks for $\ell \lesssim 200$ but no large peaks $\ell \gtrsim 600$ due to the highly damping nature of the tensor transfer function (see e.g.\cite{Pritchard:2004qp,Namba:2015gja,Thorne:2017jft}). Hence, it is difficult to reproduce these $EB$ peaks without a tuning. Although it might be possible to find a suitable oscillating spectrum of CGWs, that would not impact on our conclusion drawn below. 

\section{\label{BB result}Overproduction of $BB$ spectrum}

In this section, we consider $BB$ power spectrum which must be simultaneously produced through Eq.~\eqref{eq:EBBB}. 
Since CMB observations have measured the $BB$ power spectrum for the relevant $\ell$ range, one should seek an appropriate parameter set of the CGWs, which does not produces a too large $BB$, while reproducing $EB$. 

In the right panel of Fig.~\ref{fig:EBandBB}, 
the colored lines show the calculated $BB$ spectra induced by the CGWs for the same sets of the CGW parameters as the previous section.
The observed $BB$ data taken by SPTPol~\cite{SPT:2019nip} (blue points) and POLARBEAR~\cite{Adachi:2019mjv} (red points) are also shown with error bars. 
From this panel, one can see that the induced $BB$ spectra from the CGWs far exceed the measured amplitude and thus the corresponding parameters should be excluded.
We note that even the blue line, which is adjusted to the first peak and has the smallest height, also overproduces the $BB$ spectrum.
This result implies that it is hard for the CGWs to explain the observed $EB$ spectrum without conflicting with the observed $BB$.

\begin{figure}
    \centering
    \includegraphics[width=\linewidth]{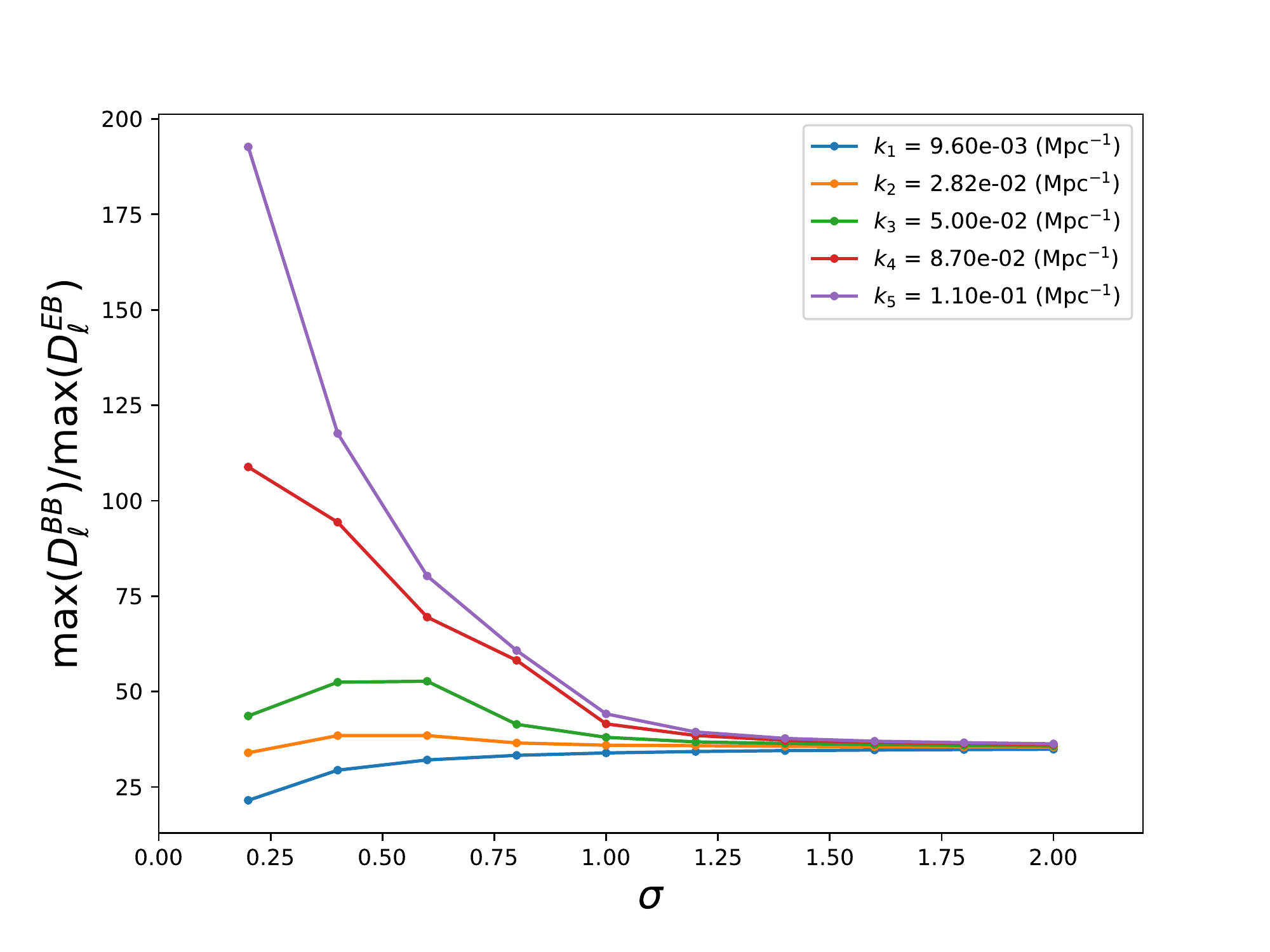}
    \caption{Ratios of maximum $D_\ell^{BB}$ to maximum $D_\ell^{EB}$ against varied width parameter $\sigma$ of $\mathcal{P}_h^R(k)$.
    The color scheme is the same as Fig.~\ref{fig:EBandBB}.
    The ratio is always larger than $21.5$, while it should be at least smaller than $2.4$ to explain the observed $EB$ without overproducing the $BB$ spectrum. 
    }
    \label{fig:BBperEB}
\end{figure}

We have fixed the width parameter $\sigma$ for the CGWs as $\sigma=0.2$ so far.
Does a different value of $\sigma$ alleviate the incompatibility between $EB$ and $BB$?
To test this possibility, we compute the ratio of the maximum value of $D_\ell^{BB}$ to that of $D_\ell^{EB}$ by increasing $\sigma$ for each contribution in Eq.~\eqref{eq:Ph_template} separately. 
As apparent from Fig.~\ref{fig:EBandBB}, since the peak height in the observed $D_\ell^{EB}$ is a few times $0.1 \, \mu \mathrm{K}^2_\mathrm{CMB}$ and the maximum value of the observed $D_\ell^{BB}$ is $0.24  \, \mu \mathrm{K}^2_\mathrm{CMB}$, this ratio $\mathrm{max}[D_\ell^{BB}]/\mathrm{max}[D_\ell^{EB}]$ should be less than $2.4$ at least.
The result is shown in Fig.~\ref{fig:BBperEB}.
Although the ratio changes depending on $k_i$, it never becomes smaller than $21.5$ for $\sigma\ge 0.2$.
For larger $\sigma$, the ratios converges to the same value around $36.5$,
because the CGW spectrum $\mathcal{P}_h^R$ becomes flatter and less dependent on its peak position $k_i$. Note that this ratio does not depend on $r_i$.
Therefore, we find that larger $\sigma$ does not mitigate the problem of the 
overproduction of $BB$ spectrum.
Even though we can make the $\mathrm{max}[D_\ell^{BB}]/\mathrm{max}[D_\ell^{EB}]$ ratio smaller with $\sigma$ smaller than $0.2$, the peaks of the $EB$ spectrum from the CGWs would be too sharp to explain the target $EB$ spectrum.

\section{\label{sec:conclusion} Summary and discussion}

Recently, the $EB$ power spectrum, which is well explained by cosmic birefringence with rotation angle $\beta\approx \SI{0.3}{deg}$, has been observed in CMB data.
In this paper, we investigated the possibility that this observed $EB$ spectrum is produced by primordial chiral gravitational waves (CGWs) instead of cosmic birefringence.
However, we found that if CGWs produced a similar $EB$ spectrum to the observed one, they would inevitably overproduce a $BB$ spectrum whose amplitude is much larger than the measured value by SPTPol and POLARBEAR.
Therefore, it is difficult to attribute the observed $EB$ spectrum to CGWs. 

To parameterize the CGW spectrum, we superposed five log-normal spectra with different peak heights and positions in Eq.~\eqref{eq:Ph_template} and analyzed it. Nonetheless, we expect that our conclusion does not depend on the detailed shape of the CGW spectrum, because the $EB$ and $BB$ spectra are linear function of $\mathcal{P}_h^R(k)$. Moreover, this CGW template enabled us to illustrate that even CGWs reproducing only one peak of the $EB$ spectrum lead to overproduction of $BB$ and should be excluded.

The reason for the difficulty of the CGW scenario can be understood as follows.
The newly observed $EB$ spectrum is smaller than the standard $EE$ spectrum mainly contributed by the scalar perturbation by a factor of $C_\ell^{EB,\mathrm{obs}}/C_\ell^{EE,\mathrm{scalar}}=\sin(4\beta)/2 \approx 10^{-2}$ for $\beta \approx 0.3^\circ$.
On the other hand, it has been known that scale-invariant primordial gravitational waves produce the $EE$ and $BB$ spectra of roughly equal size, $C_\ell^{EE,\mathrm{tens}}\simeq C_\ell^{BB,\mathrm{tens}}\simeq 10^{-4} r\, C_{\ell}^{EE,\mathrm{scalar}}$ where and hereafter we consider $\ell\approx 600$, and $r$ is the tensor-to-scalar ratio.
Thus, we expect that CGWs produce a $EB$ spectrum, $C_\ell^{EB,\mathrm{tens}}\simeq 10^{-2} r\, C_{\ell}^{EB,\mathrm{obs}}$,
and in order to reproduce the observed $EB$ spectrum, $r\simeq 10^2$ is necessary, which leads to $C_\ell^{BB,\mathrm{tens}}\simeq 10^{-2}\, C_{\ell}^{EE,\mathrm{scalar}}$.
This is incompatible with an observed fact at high $\ell$ that $C_\ell^{BB,\mathrm{obs}}\simeq C_\ell^{BB,\mathrm{lens}}\simeq 10^{-3}C_{\ell}^{EE,\mathrm{scalar}}$ with $C_\ell^{BB,\mathrm{lens}}$ the lensing $B$-mode spectrum.
Note that the above argument assumed scale-invariant CGWs and derived smaller amplitude parameter than $r_i\sim 10^{3}$ obtained in Sec.~\ref{sec:result}.
Nonetheless, it illustrates the basic reason why CGWs cause the incompatibility between the $EB$ and $BB$ spectra.


In this paper, we did not study the case with an extremely small width parameter, $\sigma < 0.2$.
The trend in Fig.~\ref{fig:BBperEB} infers that smaller $\sigma$ would increase the ratio, $\mathrm{max}[D_\ell^{BB}]/\mathrm{max}[D_\ell^{EB}]$, for the fourth (red) and fifth (purple) peaks and thus worsen the conflict between $EB$ and $BB$. Even if the ratio decreases for much smaller $\sigma$, such a spike-like $D_\ell^{EB}$ would not be responsible for the entire observed $EB$ spectrum. 
While a dedicated analysis should be done to rigorously dismiss this possibility, we expect that smaller $\sigma$ would not give a viable solution.

\begin{acknowledgments}
This work was supported in part by Japan Society for the Promotion of Science (JSPS) KAKENHI, Grants Nos.~JP18K13537 (T.F.), JP20H05854 (T.F.), JP20H01932 (S.Y.), JP20K03968 (S.Y.), JP20K14497 (Y.M.), JP19K14718 (M.S.) and JP20H05859 (M.S.). 
The authors thank the Yukawa Institute for Theoretical Physics at Kyoto University. Discussions during the YITP workshop YITP-T-21-08 on ``Upcoming CMB observations and Cosmology'' were useful to complete this work.
The authors are grateful to Yuji Chinone for suggestions on the plots of power spectra.
M.S. acknowledges the Center for Computational Astrophysics, National Astronomical Observatory of Japan, for providing the computing resources of Cray XC50.
\end{acknowledgments}


\bibliography{main}

\end{document}